# Metal Oxide-Vertical Graphene Nanosheets for 2.6 V Aqueous Electrochemical Hybrid Capacitor

Subrata Ghosh[1*], S. R. Polaki[2], Gopinath Sahoo[2], En-Mei Jin[1], M. Kamruddin[2], Jung Sang Cho[3], Sang Mun Jeong[1*]

[1] *Department of Chemical Engineering, Chungbuk National University, 1 Chungdae-ro, Seowon-Gu, Cheongju, Chungbuk 28644, Republic of Korea*
[2] *Surface and Nanoscience Division, Materials Science Group, Indira Gandhi Centre for Atomic Research- Homi Bhabha National Institute, Kalpakkam, Tamil Nadu 603102, India*
[3] *Department of Engineering Chemistry, Chungbuk National University, Chungbuk 361-763, Republic of Korea*

**Abstract**

Aqueous asymmetric electrochemical capacitor, with their high power density and superior cycle stability in comparison to conventional batteries, are presently considered as the most promising contender for energy storage. However, fabricating an electrode material and choosing a suitable aqueous electrolyte are vital in developing an electrochemical capacitor device with high charge storage capacity. Herein, we report a feasible method to synthesize $MnO_2$/Vertical graphene nanosheets (VGN) and $Fe_2O_3$/VGN as positive and negative electrodes, respectively. The surface of VGN skeleton is independently decorated with $MnO_2$ having sponge gourd-like morphology and $Fe_2O_3$ having nanorice like morphology. A schematic representation of the growth mechanism for metal oxide on VGN network is established. Both the electrodes have shown around 250 times higher charge-storage capacity than the bare VGN (0.47 $mF/cm^2$) with the specific capacitance of 118 ($MnO_2$/VGN) and 151 $mF/cm^2$ ($Fe_2O_3$/VGN). In addition to the double layer capacitance contribution, the porous interconnected vertical graphene architecture serves as a mechanical backbone for the metal oxide materials and provides multiple conducting channels for the electron transport. The fabricated asymmetric device exhibited a specific capacitance of 76 $mF/cm^2$ and energy density of 71 $\mu Wh/cm^2$ with an excellent electrochemical stability up to 12000 cycles, over a potential window of 2.6 V. The commendable performance of asymmetric electrochemical capacitor device authenticated its potential utilization for next-generation portable energy storage device.

Corresponding authors: Subrata Ghosh (subrata.ghoshk@rediffmail.com), Sang Mun Jeong (smjeong@chungbuk.ac.kr)

1. **Introduction**

With the ever-rising proliferation of portable and flexible energy storage devices, electrochemical capacitor (EC) have drawn the significant interest, owing to their higher power density and longer cyclic stability in comparison to conventional batteries.[1, 2] However, their major drawback is low energy density compared to the conventional batteries, which limits their commercialization on a large scale. The phase space available to enhance the energy density is spanned by two major elements – the electrode material and the electrolyte.[1, 2] An electrolyte with higher potential window and dielectric constant is found to enhance the energy density.[3] Since the electrode-electrolyte interactions play a major role, it is important to engineer the electrode materials, such that the increase in effective energy density is quite significant.[4]

Recently, carbon nanostructures and 2D materials have been widely studied as EC electrode materials, followed by the transition metal oxides/nitrides (TMO/N) or hydroxides and conducting polymers.[1, 5-9] However, the use of a binder, issue of restacking and agglomeration during electrode fabrication are observed to lower the effective surface area and increase the internal resistance thereby limiting the power density. Vertical graphene nanosheets (VGN) are one of the beneficial architectures of 2D materials, which has drawn a significant attention in this context. Often it is referred to carbon nanowalls, graphitic petals or few layered graphene nano-flakes. The VGN is an interconnected open porous network of vertically standing few-layer graphene sheets.[10] Owing to its compatibility with electrode materials and superiority compared to other conventional electrodes in all aspects, yet its specific capacitance in the range of microfarad/cm$^2$ to few millifarad/cm$^2$ only.[4, 11] Apart from the double layer capacitance contribution, VGN also holds a promise to serve as a mechanical backbone of electrochemical capacitor electrode materials as well as provides conductive channels and multiple pathways for the electron and ion transport.[12] On the other hand, transition metal oxides (TMO) are often encouraging due to their higher theoretical capacitance even ten to hundred times higher than that of the carbon materials.[8, 13] To overcome the flaws of pseudocapacitors like poor electrical conductivity and inferior electrochemical stability of TMO, an amalgamation of TMOs with carbon matrix is the traditional roadmap for the researchers. Noteworthy that the vertical oriented structure is more beneficial as energy storage electrode compared to its horizontal counterpart owing to higher accessible surface area, larger edge density and multiple channels. Therefore, metal oxide coating on the vertically oriented carbon skeleton is an advantageous strategy to obtain high-performance energy storage device.

Furthermore, widening the potential window of aqueous electrolyte and in-turn increasing the energy density via the relation, $E = 0.5CV^2$ is a key challenge for the energy research community. In this context, chemical activation of the electrode materials, use of novel electrolyte, fabrication of asymmetric electrochemical supercapacitor (ASC) device are the promising approaches to enhance the electrolyte-electrode interaction, widening potential window and hence the energy density.[14-17] However, the maximum reported potential window for aqueous ASC is limited within 1.6 to 2.0 V [18-22] and countable reports exhibited an aqueous ASC device with a potential window higher than 2V. [14, 23, 24] In order to obtain ASC device with highest possible potential window, one should choose a proper pair of TMOs since the potential window also depends on the difference in work-function between the electrodes. Amidst TMOs, $MnO_2$ and $Fe_2O_3$ nanostructures are more popular, due to their higher capacity, abundances, low cost, non-toxicity and environment-friendly.[19, 21] Inspite of the significant progress in architecting $MnO_2$ and $Fe_2O_3$ electrode that has been accomplished in SC performance, the complex fabrication process, cost, agglomeration and limited electrochemical performance hindered scaling-up its production. Hence, our scientific research has been focused towards designing a self-standing TMO- vertical graphene hybrid structure via an easy and facile method to construct a high voltage ASC device.

Herein, we constructed an aqueous based asymmetric Electrochemical capacitor device comprised of binder-free and additive free, $MnO_2$/VGN and $Fe_2O_3$/VGN nanoarchitectures as the positive and negative electrode, respectively. We propose a facile and scalable method to prepare the sponge gourd-like $MnO_2$ and nanorice-like $Fe_2O_3$ structures onto each sheet of vertical graphene with preserving vertical orientation and open network, without the aid of binder and the conductive agent. The fabricated device ensured its ability to operate with the high cell voltage of 2.6 V in 1M sodium perchlorate electrolyte ($NaClO_4$). In combination of open network and vertical orientation, metal oxide-VGN provides efficient accessible surface area and conductive pathways for the electrolyte ions. The advantages of such devices include an enhanced capacitance of 76 mF/cm$^2$, excellent stability and higher energy density.

1. **Experimental method**
2.1. Electrode Fabrication
*2.1.1 Vertical graphene growth*
The VGN was grown using plasma enhanced chemical vapor deposition on the graphite paper substrate. The details of the growth process were elaborated in our previous report.[25] The VGN was used as a template for the growth of $MnO_2$ and $Fe_2O_3$ through the wet-chemistry and hydrothermal method, respectively, which were later used as the positive and negative electrode materials of the asymmetric device. In order to fabricate the metal oxide/VGN composites, firstly the as-prepared VGN were subjected to KOH treatment at 100°C for 30 min to achieve better graphitic quality, wetting property and hence its Electrochemical capacitor performance. The contact angle measurements revealed the reduction in wetting angle from $110^0$ to $76^0$, which was much less than our previous reported value.[15]

*2.1.2. Metal oxide-Vertical graphene growth*
The fabrication of $MnO_2$/VGN electrode was achieved by dip-coating followed by annealing at 400 $^0$C. Briefly, the precursor solution of $MnO_2$ nanostructure was prepared by dissolving $KMnO_4$ in deionized water under constant magnetic stirring. An activated VGN was dipped into $KMnO_4$ solution and consecutively subjected to keep the solution at 80 $^0$C for 30 min. Subsequently after the growth, the electrode was rinsed in DI water and ethanol. Finally, the samples were loaded into the furnace and allowed them to anneal for 2h at 400 $^0$C in Ar environment (Scheme 1). The growth of $MnO_2$ nanostructures on VGN occurred via the following reaction:

$$KMnO_4(aq.) + C(s) + H_2O \rightarrow MnO_2(s) + CO_3^{2-} + HCO_3^{-} \quad (1)$$

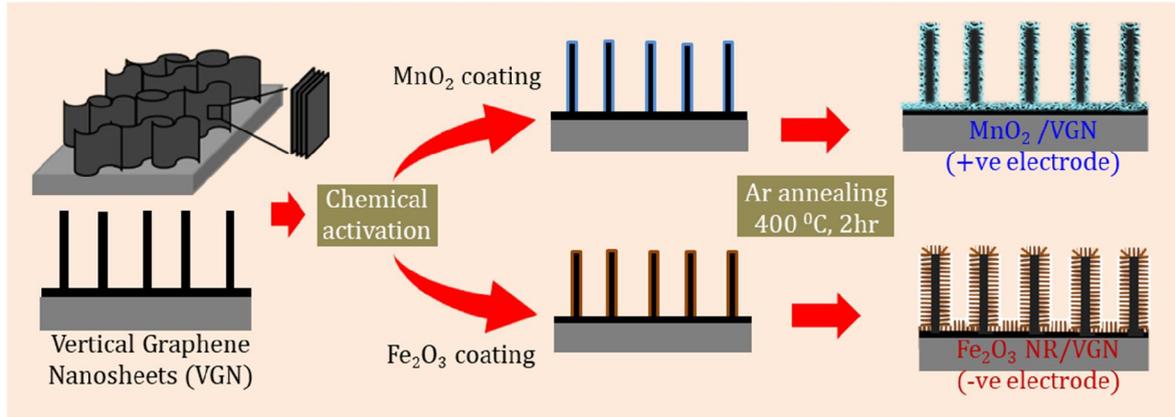

*Scheme 1: Schematic of the fabrication process of positive and negative electrodes based on metal oxide-vertical graphene nanosheets*

The bare $Fe_2O_3$ nanorice were fabricated by hydrothermal methods to compare its performance with that of $Fe_2O_3$/VGN.[26] The graphite paper and activated VGN were immersed into $FeCl_3 \cdot 6H_2O$ ethanol/aqueous precursor solution to prepare a bare $Fe_2O_3$ nanorices and $Fe_2O_3$/VGN, respectively. The volume ratio of ethanol/aqueous was maintained at 1:1. Following which, the mixture was hydrothermally treated for 2h at 120 $^0$C using the Teflon-lined autoclave. Samples were taken out, washed by distilled water and ethanol for several times. The VGN was uniformly coated with a yellow product of FeOOH. Before subjected to anneal, the composites were freeze-dried. Then the samples were loaded into the furnace and annealed for 2h at 400 $^0$C in Ar environment to achieve $Fe_2O_3$ nanorices structure (Scheme 1). The chemical reaction involved in the hydrothermal and subsequent annealing steps are

$$FeCl_3(aq.) \rightarrow Fe^{3+}(aq.) + 3(OH)^- \rightarrow FeOOH \rightarrow \alpha Fe_2O_3 + 3H_2O \quad (2)$$

2.2. Morphological and structural characterization

The morphology of as-grown and metal oxide decorated VGN was inspected using Field Emission Scanning Electron Microscope (FESEM, Zeiss, ULTRA PRUS). The wettability of all the studied samples was measured by water contact angle measurement using the sessile drop method (GSM, Surfacetech Co., Ltd.). The structural confirmation of all samples was investigated by Raman Spectroscopy (in-Via Renishaw) using 532 nm laser. The chemical composition and oxidation states were evaluated using X-Ray Photoelectron Spectroscopy ((XPS, Thermo Scientific, K-Alpha) with Al K$_\alpha$ (1486.7 eV) X-ray radiation.

2.3. Electrochemical investigation

The electrochemical performance of each individual electrode was evaluated by cyclic voltammetry at different scan rates and charge-discharge at different current densities. The specific capacitance of electrodes ($C_s$) in F/cm$^2$ were estimated by the relations

$$C_s = \frac{\int I dV}{2 \times A \times v \times \Delta V} \quad (3)$$

$$C_s = \frac{I \times t_d}{A \times \Delta V} \quad (4)$$

where $\int IdV$, $A$, $v$, $\Delta V$ and $t_d$ are the area under the CV profile, exposed area of the electrode in electrolyte, scan rate, the operating potential window and discharge time, respectively. The individual electrodes were tested in the three-electrode configuration in which graphite rod and Ag/AgCl (Saturated in 3M KCl) were used as the counter and reference electrodes, respectively. The supercapacitive behavior of asymmetric device was evaluated in the two-electrode configuration. To assemble the ASC device, NaClO$_4$ soaked cellulose paper as separator-cum-electrolyte was sandwiched between the electrodes. Autolab electrochemical workstation (AUT84455, Netherland) was employed for the above investigations. The specific energy density (Wh/cm$^2$) and power density (W/cm$^2$) of ASC device were estimated from using the following relations:

$$E = {C_s \times \Delta V^2}/{2 \times 3600} \quad (5)$$

and P= $E \times 3600/t_d$ (6)

Where E and P are the specific energy density and power density, respectively.

## 3. Result and Discussion

*3.1 Morphology and characterization of MnO$_2$/VGN*

A well-controlled MnO$_2$ structure on the VGN was obtained according to Scheme 1 by varying the molar concentration of KMnO$_4$ precursor from 2 to 20 mM. The well-controlled and uniform MnO$_2$/VGN was achieved only with the 10 mM KMnO$_4$ solution, where the open gap between the vertical nanosheets was preserved. The growth rate of MnO$_2$ nanostructure can be controlled by immersion time and deposition temperature.[27] To compare the growth of MnO$_2$ on VGN, bare MnO$_2$ nanostructures was also grown on graphite paper. It has seen that spounge-gourd like structure of MnO$_2$ on VGN whereas non-uniform coating of MnO$_2$ is observed on graphite paper (Fig. 1(a)). The inset at the left-bottom of Fig. 1(a) demonstrates the porous morphology of VGN, where the vertically standing nanosheets are interconnected to each other and forms a porous interconnected network. The average inter-sheet spacing and thickness of each graphene sheet of VGN is measured by imageJ analysis and found to be 182 (±96) nm and 3-10 nm, respectively. The sponge gourd-like uniform and homogeneous coating of MnO$_2$ over the skeleton of VGN, without filling the intersheet-spacing is ensured from the Fig. 1(a). The average inter-sheet spacing and thickness of each sheet varied from 56 to 141 nm and 54 (±12) nm, respectively after the MnO$_2$ decoration. In general, the inter-sheet spacing is considered as ion-reservoir and each vertical sheets serves as nanoelectrode.[4] This kind of morphology is beneficial for efficient ion-transport and effective accessibility for the full interior surface, which in-turn enhance the electrochemical performance. The coating of MnO$_2$ onto VGN structure is confirmed by the Raman spectra and XPS analysis. The inset of Fig. 1b illustrates, the presence of weak band centered around 660 cm$^{-1}$ corresponding to the Mn-O symmetric stretching vibration of MnO$_6$ octahedra [12, 28]. The Raman spectra of MnO$_2$/VGN also consist of D" (at 1100 cm$^{-1}$), D (at 1353 cm$^{-1}$), G (at 1586 cm$^{-1}$), D' (1621 cm$^{-1}$), D+D" (2464 cm$^{-1}$), G' (2702 cm$^{-1}$), D+D' (2946 cm$^{-1}$) and 2D' (at 3237 cm$^{-1}$) peak. These peaks are characteristic Raman peaks of the VGN.[29] The presence of $MnO_2$ is also confirmed from XPS result of $MnO_2$/VGN (Fig. 1c). The survey spectrum is the clear evidence for the presence of C, O and Mn. Figure 1d. shows the deconvoluted C1s spectrum which includes $sp^2$-C at 284.4 eV, $sp^3$-C at 284.9, oxygenated C in C-O at 285.8 eV, C=O at 286.5 eV, -COO at 287.5, vacancy defect at 283.6 and π-π* satellite peak at 292 eV.[30] These oxygenated functional groups can enhance the electrode wettability with the electrolyte and served as the bridging element for $MnO_2$. The high-resolution Mn2p peak, as shown in Fig. 1e, exhibited prominent $Mn2p_{3/2}$ and $Mn2p_{1/2}$ peaks with separation of 11.9 eV.[31, 32] This result confirms the +4 oxidation state of Mn.[24] The deconvoluted high-resolution O1s spectra shows three distinct peaks at 526.7, 531.1 and 532.3 eV and are assigned as Mn-O-Mn, Mn-O-C and Mn-O-H, respectively (Fig. 1f).[33] Wettability of the electrode materials is another influential factor to have better supercapacitive properties.[34] Noteworthy that, water is immediately spread once it drops onto the $MnO_2$/VGN surface. Whereas, the contact angle of $MnO_2$ surface was found to be in the range 26 to 77.9$^0$ in the literature.[35] The observed fact indicates that $MnO_2$/VGN is superhydrophilic in nature which may lend the excellent Electrochemical capacitor performance. This hydrophilicity has arisen from the interaction between Mn-O-H group.[36]

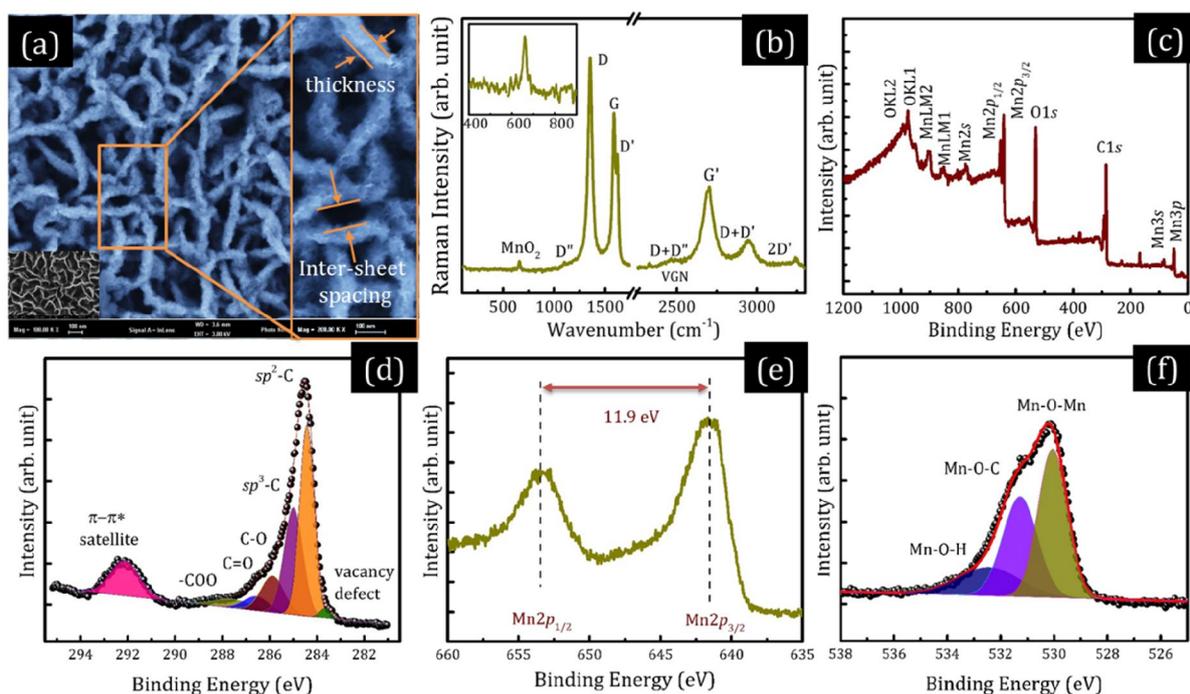

*Figure 1: (a) Electron micrograph with pristine VGN in insight, (b) Raman spectra, (c) X-ray photoelectron survey spectra; (d) high-resolution C1s spectra, (e) high-resolution Mn2p spectra and (f) high-resolution O1s spectra of $MnO_2$/VGN*

*3.2 Morphology and characterization of $Fe_2O_3$/VGN*
Figure 2(a) displays the scanning electron micrographs of $Fe_2O_3$ nanorice decorated VGN. It is clear that $Fe_2O_3$ nanorice is grown on the surface of VGN as well as on its edges. However, it is noteworthy to mention that the porous networked structure of VGN is well-maintained even after controlled $Fe_2O_3$ nanorice coating. The average length and diameter of nanorice are 30 to 74 nm and 18 (±2) nm, respectively. The average inter-sheet spacing and thickness of each sheet varied from 74 to 250 and 74 (±19) nm after $Fe_2O_3$ nanorice growth over VGN, respectively. To compare the growth of $Fe_2O_3$ nanorice on VGN, the bare $Fe_2O_3$ nanorice was grown on graphite paper. An agglomerated and non-uniform $Fe_2O_3$ nanorices was observed on graphite paper. Whereas, non-agglomerated and uniform $Fe_2O_3$ nanorices are obtained while they are grown on VGN. This fact ensured the advantage of VGN as a mechanical platform to nurture uniform and homogeneous growth of hetero-nanostructures. However, the process parameters were well controlled and tuned to get $Fe_2O_3$ nanorices-VGN composites, while preserving the VGN geometry. The $Fe_2O_3$/VGN composites were further evaluated by Raman spectroscopy and XPS. The Raman spectra of $Fe_2O_3$/VGN shows the characteristic peaks of both VGN and $Fe_2O_3$ (Fig. 2b). The inset of Fig. 2(b) shows the Raman spectra of $Fe_2O_3$. Two strong and broad peaks at 346 ($T_1$) and 718 cm$^{-1}$ ($A_1$) have clearly appeared along with a very weak peak at 500 cm$^{-1}$ (E) in the Fig. 2(b). However, the magnon mode at 1300 cm$^{-1}$ of $Fe_2O_3$ is overlapped with the D-peak of VGN.[37] Furthermore, the XPS measurement is carried out to evaluate the composition of the $Fe_2O_3$/VGN structure. The XPS survey spectrum exhibited C1$s$ peak at 285 eV, and O1$s$ peak at 530 eV accompanied with features of $Fe_2O_3$ (Fig. 2c-f). The high-resolution Fe2$p$ spectra consists of Fe2$p_{3/2}$ peak at 710.9 eV and Fe2$p_{1/2}$ peak at 724.6 eV associated with their respective distinguishable satellite peaks at 718.9 and 732.6 eV (Fig. 2e). In conjunction, the binding energy difference between Fe2$p_{3/2}$ and its satellite peaks is around 8 eV.[38] The O1$s$ spectrum of $Fe_2O_3$/VGN was deconvoluted into three components. The peaks at 529.9, 531.3 and 533 eV in O1$s$ spectrum are attributed to the Fe-O, Fe-O-C and C-O bond, respectively (Fig. 2f).[39] Hence, the features observed in both Raman and XP spectra confirms the successful decoration of $Fe_2O_3$-nanorice structure onto the VGN matrix. Like $MnO_2$/VGN, the $Fe_2O_3$/VGN is also found to be super-hydrophilic in nature since water spread immediately once it drops on the surface.

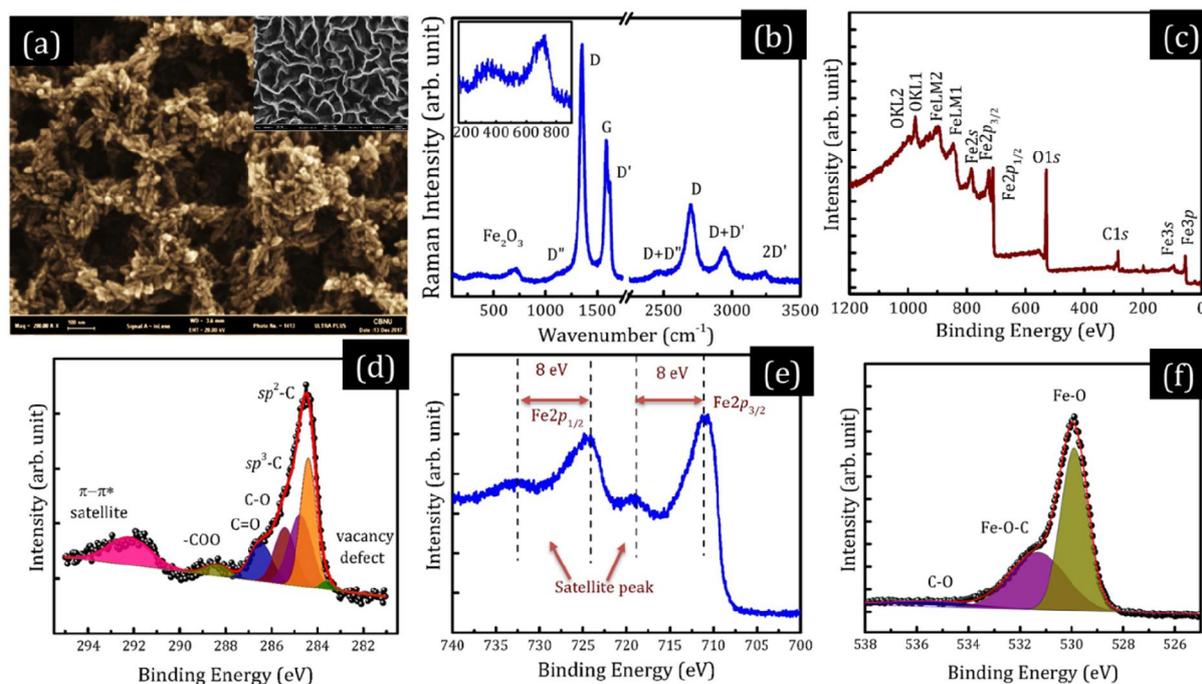

*Figure 2: (a) Micrograph with inset of as-prepared VGN, (b) Raman spectra, (c) X-ray photoelectron survey spectra, (d) high-resolution C1s spectra, (e) high-resolution Fe2p spectra and (f) high-resolution O1s spectra of $Fe_2O_3$/VGN.*

3.3 *Plausible formation mechanism of metal oxides on VGN*

Based on the above investigation, we propose a step-wise formation mechanism of metal oxide/VGN as follows (Fig. 3). The growth mechanism of the VGN is already discussed in our previous reports.[25, 40] The plasma-based growth of VGN is responsible due to the in-built electric field, competition between growth/etching process of carbon radicals, plasma density gradient towards the substrate and released strain at the grain boundary of nanoislands. Finally, the VGN formation takes place with partially terminated H-edges.[25] This terminated H-edges are replaced by more oxygenated function groups while they undergo KOH activation. Since KOH activation of the VGN surface leads to the higher amount of oxygen functional groups.[15] This oxygenated functional group serves as the nucleation center for the TMO coating. In the next step of TMO coatings, oxygen group of metal oxides are reacting with active edges of VGN through replacement of H by metal ions or attaching with oxygen presence on the carbon surface. Finally, M-O-C formation occurs which leads to the successful coating of nanostructured metal oxides on VGN. The difference in morphology of both metal oxides is due to the procedure applied to grow individual metal oxide nanostructures.

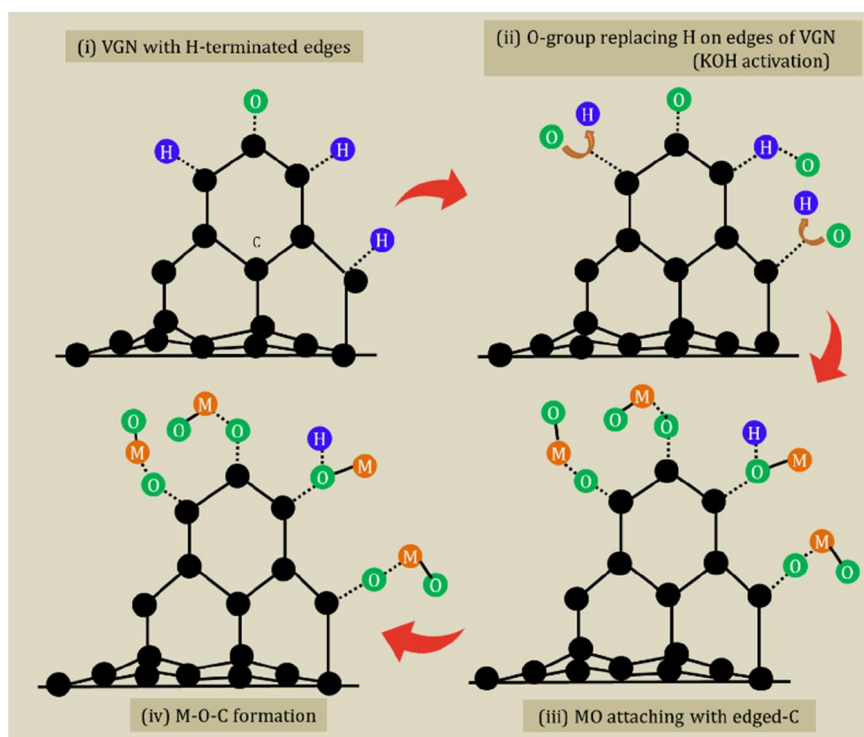

*Figure 3: Plausible formation scheme of vertical metal oxide-graphene composite (dotted lines corresponds to physical interactions among atoms)*

3.4 *Electrochemical performance of MnO$_2$/VGN*

Prior to the electrochemical performance of vertical metal oxide-graphene composite, the supercapacitive properties of VGN was evaluated. The estimated specific capacitance of VGN is found to 0.47 mF/cm$^2$ at scan rate of 100 mV/s. Moreover, with the advantages of edges, high surface area and open network, the vertical structures are more efficient in charge-storage compared to planar graphitic structures.[40] A near-rectangular and scan rate independent shape of CV assures the ideal capacitive behavior of VGN in both sides of the potential window. The electrochemical performance of MnO$_2$/VGN in the three-electrode system is displayed in the Fig. 4. The quasi-rectangular, reversible and shape-preserved cyclic voltammogram with scan rate is observed from Fig. 4a. The absence of oxidation/reduction peak in CV ensures the faster reversible redox reaction of MnO$_2$. The comparative CV of MnO$_2$/VGN with VGN and bare MnO$_2$ ensures the higher charge storage capacity of MnO$_2$/VGN composite (Fig. 4b). In addition, it has shown that the bare MnO$_2$ and MnO$_2$/VGN can be safely operated upto 1.3 V in the positive window of potential (Fig. 4a). This fact is in well-agreement with the existing report.[33] In addition, the increased area under the CV curve of bare MnO$_2$ is observed when the potential window is increased from (0-1.0V) to (0-1.3 V). An enhanced CV area and therefore specific capacitance with respect to the applied potential window are due to the increased dielectric constant of electrolyte as well as reduced charge-separation distance.[36] The maximum specific capacitance of MnO$_2$/VGN is found to be 118 mF/cm$^2$ at 10 mV/s and maintained 35 mF/cm$^2$ while scan rate is increased by twenty times (Fig. 4c). The reduced specific capacitance of MnO$_2$/VGN at higher scan rate is due to the inability of complete redox transitions.[36] Even, the charge-storage performance of MnO$_2$/VGN can be further enhanced by varying deposition time and temperature in KMnO$_4$ solution.[27] However, the obtained value is found to be higher than that of MnO$_2$ coated VGN (5.6 mF/cm$^2$ at 100 mV/s),[12] Mn-Mo oxide – CNTs composite (31 mF/cm$^2$ at 2 mV/s),[41] MnO$_2$ based composites (79.15 mF/cm$^2$ at 0.4 mA/cm$^2$),[42] MnO$_2$ coated carbon nanoparticles (109 mF/cm$^2$ at 5 mV/s)[43]. In order to evaluate the charge-storage behavior, the current density with respect to scan rate is plotted in Fig. 4d. It is observed that the current density *vs* scan rate plot follows the power law relation: $i = av^b$. The value of b is determining factors for charge storage mechanism, wherein b=1 corresponds to pure capacitive response and 0.5 refers to semi-infinite diffusion-controlled charge-storage. From the linear fitting, the slope is estimated to be 0.56, indicating the semi-infinite diffusion-controlled charge-storage mechanism of the MnO$_2$/VGN electrode (Fig. 4d). The charge-storage mechanism of MnO$_2$/VGN can be ascribed to the fast intercalation/de-intercalation of Na$^+$ ions into/out of the matrix and surface adsorption of Na$^+$ via following reactions, respectively [36]

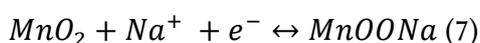
$$MnO_2 + Na^+ + e^- \leftrightarrow MnOONa \quad (7)$$

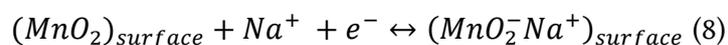
$$(MnO_2)_{surface} + Na^+ + e^- \leftrightarrow (MnO_2^- Na^+)_{surface} \quad (8)$$

The charge-discharge (C-D) profile at various current densities, as shown in Fig. 4e, confirmed the ideal capacitive behavior and excellent rate performance. Figure 4f shows the capacitance retention of MnO$_2$/VGN composite for 2000 C-D cycles. The excellent charge-storage behavior of the composite is attributed to the double-layer contribution from VGN and pseudocapacitance of MnO$_2$ along with the excellent electrical conductivity of the VGN as well as the hydrophilicity of the MnO$_2$/VGN.

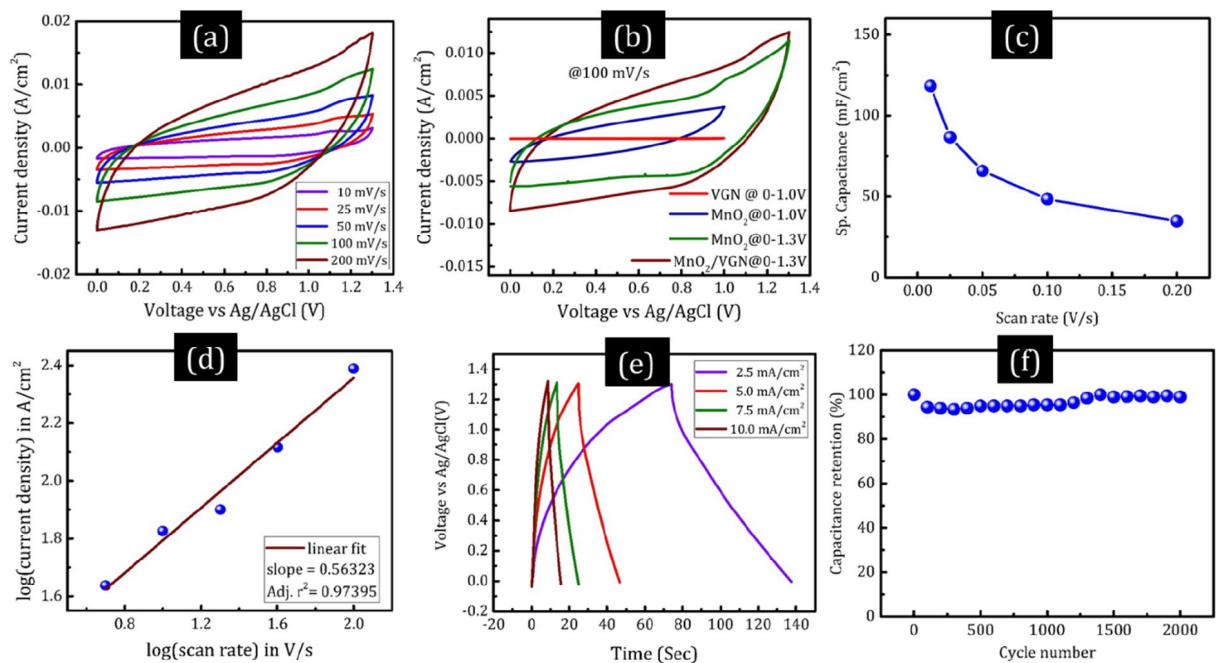

*Figure 4: (a) cyclic voltammogram at different scan rate, (b) comparative CV with respect to bare VGN and MnO₂, (c) specific capacitance with respect to scan rate, (d) current vs scan rate profile, (e) charge-discharge profile and (f) Capacitance retention with charge-discharge cycle for MnO$_2$/VGN electrode*

### 3.5 Electrochemical performance of Fe$_2$O$_3$/VGN

The electrochemical performance of Fe$_2$O$_3$/VGN is investigated, while it was tested as the negative electrode (Fig. 5). A well-defined cathodic peak at -1.06 V and anodic peak at around -0.58 V are observed and they are attributed to the reversible oxidation and reduction process (Fig. 5a and S5). Along with that, a weak cathodic peak at -0.26V and anodic peak at -0.67V are also emerged when the window is increased from 1V to 1.35 V (Fig. 5b). These weak peaks are prominent for Fe$_2$O$_3$/VGN in a lower scan rate and shown in Fig. S6. Other notable feature is the reversible redox peaks are (i) shifted with respect to the scan rate for Fe$_2$O$_3$/VGN (represented by dashed line in Fig. 5a) as well as (ii) stronger and shifted when Fe$_2$O$_3$ nanorices are decorated on VGN (Fig. S6). This fact ensures that the Fe$_2$O$_3$ nanorices are well-grafted on VGN with optimized density. The possible redox reaction here is as following:[44]

$$Fe_2O_3 + 3H_2O + 2e^- \leftrightarrow 2Fe(OH)_2 + 2OH^- \quad (9)$$

$$Fe_2O_3 + 2Na^+ + 2e^- \leftrightarrow Na_2Fe_2O_3 \quad (10)$$

$$FeOOH + H_2O + e^- \leftrightarrow Fe(OH)_2 + OH^- \quad (11)$$

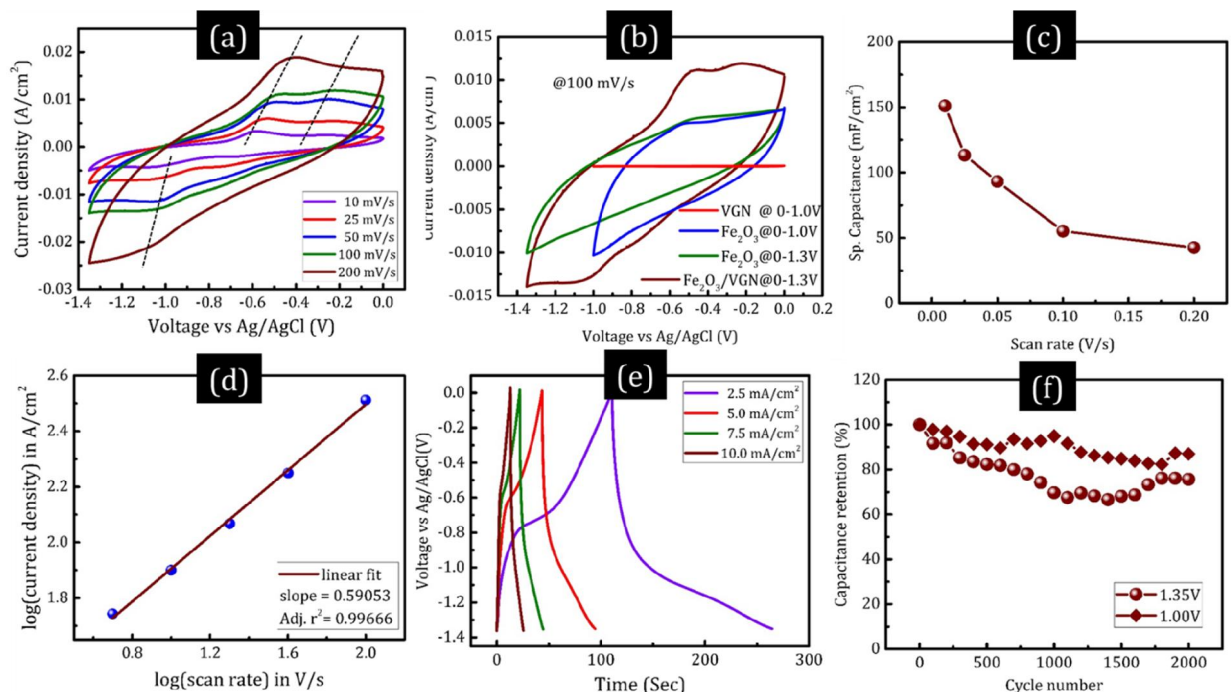

*Figure 5: (a) cyclic voltammogram at different scan rate, (b) comparative CV with respect to bare VGN and Fe$_2$O$_3$, (c) specific capacitance with respect to scan rate, (d) current vs scan rate profile, (e) charge-discharge profile, (f) Capacitance retention with charge-discharge cycle for Fe$_2$O$_3$/VGN electrode*

The effect of Fe$_2$O$_3$ nanorice decoration onto VGN structure and thus higher charge-storage capacity is clearly observed from the comparative CV with regards to VGN and bare Fe$_2$O$_3$ nanorices (Fig. 5b). The above result also indicates the widened hydrogen evolution overpotential of Fe$_2$O$_3$/VGN and hence ability to operate the composite electrode over the potential window of 0 to -1.35V without any hydrogen evolution.[45] The areal specific capacitance of Fe$_2$O$_3$/VGN is estimated with respect to the scan rate and plotted in Fig. 5c. The highest area-specific capacitance for the VGN is found to be 151.11 mF/cm$^2$ at 10 mV/s and maintained to 42.68 mF/cm$^2$ at 200 mV/s. The estimated capacitance of Fe$_2$O$_3$/VGN is much higher than that of graphene-wrapped Fe$_2$O$_3$ nanowire network (3.3 mF/cm$^2$ at 10 mV/s),[44] α-Fe$_2$O$_3$@PANI (103 mF/cm$^2$ at 0.5 mA/cm$^2$),[46] Fe$_2$O$_3$ hollow nanorod (around 80 mF/cm$^2$ at 10 mV/s),[47] and Fe$_2$O$_3$ nanotube annealed at 200° (~20 mF/cm$^2$ at 100 mV/s), 400° (~70 mF/cm$^2$ at 100 mV/s), 500° (~30 mF/cm$^2$ at 100 mV/s) and 600° C (~15 mF/cm$^2$ at 100 mV/s)[48]. In our case, the Fe$_2$O$_3$/VGN was annealed at 400 °C. The areal capacitance of Fe$_2$O$_3$/VGN can be improved further by controlling the aspect ratio of nanorices and annealing temperature.[48, 49] The presence of well-defined redox peaks in CV is a clear indication of the pseudocapacitive charge storage mechanism. However, to quantify its contribution, we have estimated the b-value and is found to be 0.59 as shown in Fig. 5d. Hence, like MnO$_2$/VGN, the semi-infinite diffusion-controlled charge storage behavior is also confirmed for Fe$_2$O$_3$/VGN. The charge-discharge profile of Fe$_2$O$_3$/VGN, is displayed in Fig. 5e, exhibited a typical pseudocapacitive features and in good-agreement with CV profile. The higher discharge-time of the Fe$_2$O$_3$/VGN compared to its charging time is observed from its charge-discharge profile. Even, similar behavior is also observed in the CV profile of this structure, while it scanned in more negative range. This feature is attributed to the reduction of $Fe_2O_3$ to metallic Fe.[24] The $Fe_2O_3$/VGN structure showed 75.7 % capacitance retention in the potential window of 1.35V after 2000 charge-discharge cycles, whereas 87% retention is observed when it is scanned within 1V window (Fig. 1f). The 14.3% loss in capacitance retention after 2000 cycles is ascribed to the loss of active material due to volume change during redox reactions at higher window.[44] However, this loss is less compared to the bare $Fe_2O_3$, due to good bonding with the VGN though bridging oxygen functional group.

3.6 *Electrochemical performance of Asymmetric Electrochemical capacitor*
Considering a good charge-storage performance of $MnO_2$/VGN as the positive electrode and $Fe_2O_3$/VGN as negative electrode, the ASC device was fabricated by sandwiching $NaClO_4$ soaked filter paper between them, which acts as a separator-cum-electrolyte reservoir. Figure 6a shows the schematic of ASC device fabricated. Prior to the testing of the device, charge balance between the electrodes is maintained by the area ratio of electrodes via the following relationship:[47]

$$\frac{A_+}{A_-} = \frac{C_- \times V_-}{C_+ \times V_+} \qquad (12)$$

where $A_{+(-)}$, $C_{+(-)}$ and $V_{+(-)}$ are the area, specific capacitance and the potential window of positive(negative) electrode, respectively. Thus, the calculated area ratio between the electrode is 1.32:1 for our case. The CV profile of individual electrode along with the ASC device is displayed in Fig. 6b, while scanned at 100 mV/s rate in three-electrode configuration. A series of CV tests with different potential window are carried out in order to achieve the best operating voltage of ASC and shown in Fig. 6c. The scan rate dependent CV result, as indicated from Fig. 6d, reveals an ideal supercapacitive behavior. As the charge balance is one of the critical parameters for device performance, we measured the charge balance directly with the aid of GPES software. Around 90% charge-balance is observed from the experimental data (Fig. 6e). This result confirms successful optimization of area ratio for both electrodes. However, slight reduction in charge-balance is may be due to the water-splitting from the metal oxide and uncovered graphene surface.[50] Similar behavior is also observed from the charge-discharge performance of the device (Fig. 6f-g). The maximum specific capacitance of ASC device is estimated to be 76 mF/cm$^2$ at 2.5 mA/cm$^2$ with 50% capacitance retention while the current density is increased by four times. Indeed, the capacitance obtained in our study is found to be much higher than that of the $MoS_2$/graphene symmetric cell (11 mF/cm$^2$ at 5 mV/s),[51] Fiber-based $MnO_2$/CNT/polyimide ASC (59.5 mF/cm$^2$ at 0.74 mA/cm$^2$),[52] PANI/WO3 ASC (23.2 mF/cm$^2$ at 0.2 mA/cm$^2$)[53]. The estimated energy density is calculated to be 71 μWh/cm$^2$ (power density = 3.25 mW/cm$^2$), which is around 22 times higher than that of the device while operated in 1V (Fig. 6h). Furthermore, an excellent cyclic performance of the ASC device is confirmed as it is tested for 12000 charge-discharge cycles with a capacitance retention of 79% (Fig. 6i). The inset of Fig. 6i is the representation of a few charge-discharge cycles.

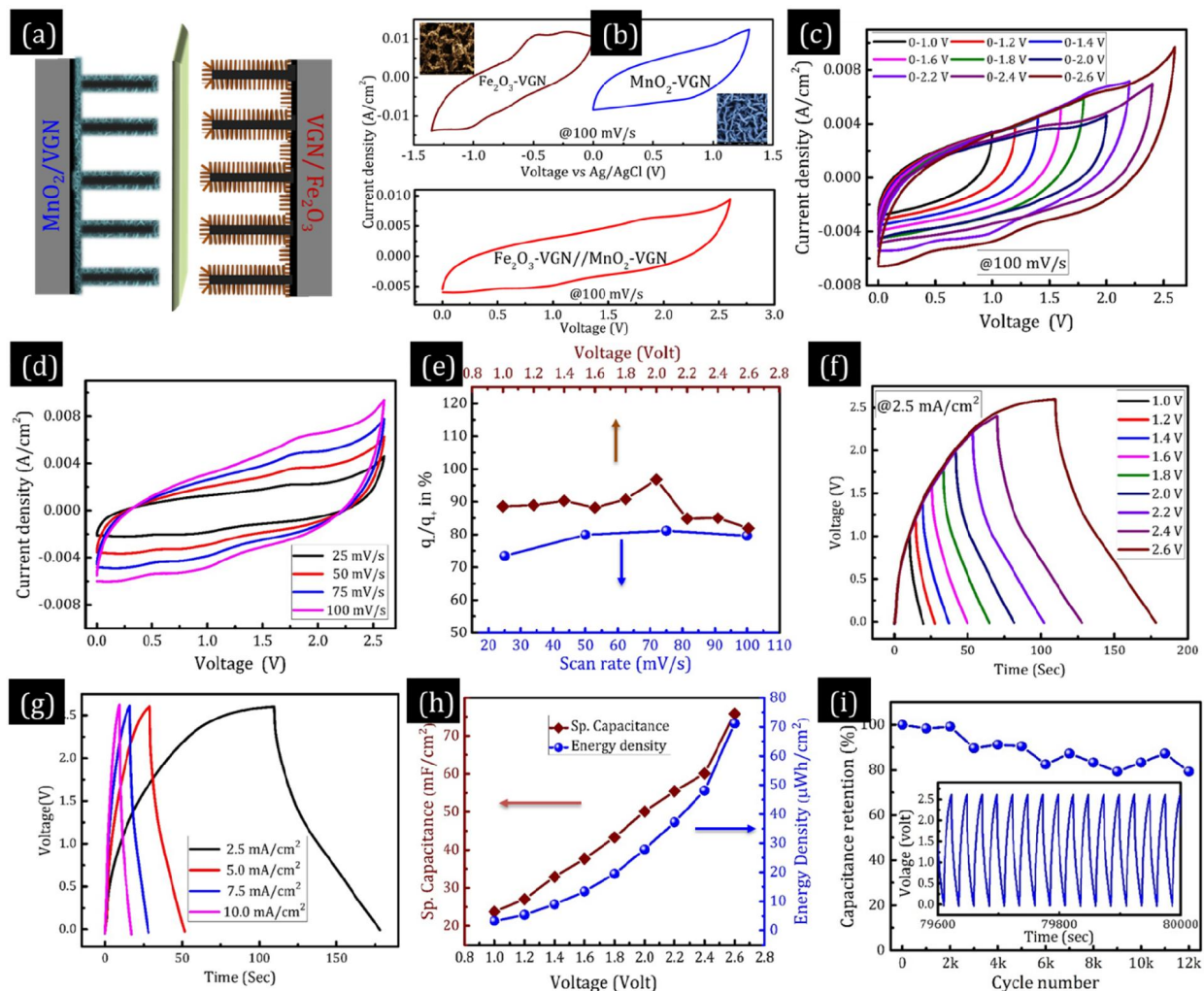

*Figure 6*: (a) Schematic of Asymmetric Electrochemical capacitor (ASC) device fabrication, (b) CV of individual electrodes at top and ASC at bottom; CV of ASC at different (c) potential window, and (d) scan rates; (e) Charge balance with respect to scan rate at constant window and applied voltage at constant scan rate; Charge-discharge profile at different (f) potential window and (g) current densities; (h) Specific capacitance and energy density of ASC with respect to the applied potential, (i) Cyclic stability and inset represents the intermediate cycle performance

The above result suggests the ability of the ASC device to operate at a stable cell voltage of 2.65 V. From the Fig. 6(c) it is clear that the stable and safe cell voltage can be extended upto 2.6 V without oxygen evaluation in the $NaClO_4$ electrolyte. The oxygen evolution is found to be started from 2.7 V. Hence, the ASC device is operated within a stable potential window of 2.6 V for further study and tested with different scan rates. The stable potential window might be caused from a synergistic effect of both positive and negative electrodes, modified work function of active materials via chemisorption of electrolyte ions as well as extended work function difference between them (Fig. 7a).[54, 55] Since water dissociation is limited kinetically via hydrogen/oxygen evolution on TMO surface, the stable operating voltage extended over the dissociation energy of aqueous electrolyte.[56] Only countable studies are reported on the ASC device with such high potential window of 2.6 V (Table 1).

*Table 1: Summary of highest potential window of aqueous asymmetric Electrochemical capacitor (CNF: carbon nanofiber; N-CNW: nitrogen doped carbon nanowire; rGOA: reduced graphene oxide aerogel)*

| Electrode | | Electrolyte | Cell voltage | Ref. |
|---|---|---|---|---|
| positive | negative | | | |
| PANI | $WO_3$ | $PVA/H_2SO_4$ | 1.4 V | [53] |
| $MnO_2$ | Graphene | PVA/LiCl | 1.5 V | [32] |
| $MnO_2$/rGOA | $Fe_2O_3$/rGOA | 0.5M $Na_2SO_4$ | 1.8 V | [57] |
| $MnO_2$/graphene | Activated CNF | 1 M $Na_2SO_4$ | 1.8 V | [58] |
| $MnO_2$/CNT | $Fe_2O_3$/CF | $PVA/LiClO_4$ | 2V | [24] |
| AgNR/rGOA | PAniNT/rGOA | $PVA/H_2SO_4$ | 1.8 V | [22] |
| $Fe_2O_3$/ N-CNW | $MnO_2$/N-CNW | PVA/LiCl | 1.6 V | [20] |
| Hydrogenated $TiO_2$ NW | $MnO_2$ | 0.5M $Na_2SO_4$ | 2.4 V | [59] |
| Fiber-based $MnO_2$/CNT | Polyimide/ CNT | $CMC/Na_2SO_4$ gel | 2.1 V | [52] |
| $Na_{0.25}MnO_2$ | porous carbon | 1M $Na_2SO_4$ | 2.7V | [23] |
| **$MnO_2$/VGN** | **$Fe_2O_3$/VGN** | **1M $NaClO_4$** | **2.6 V** | **This work** |

For further understanding the charge-storage behavior of ASC device, an impedance spectroscopy was performed in the frequency range of 100 kHz to 10 mHz with an ac perturbation of 10 mV (Fig. 7b). The impedance spectra consist of semicircular arc at high-frequency region accompanied with the steeper line at lower-frequency region. The measured impedance spectra are fitted with a suitable equivalent circuit via ZSimpWin software, which is provided in the inset of Fig. 7b. The $R_s$, CPE, $R_{ct}$, W and $C_p$ in the equivalent circuit represents equivalent series resistance, constant phase element, charge-transfer resistance, Warburg resistance and pseudocapacitance, respectively. The Warburg resistance, slope of 45° in the curve, is arisen due to the frequency dependent ion transport from electrolyte to electrode surface. The $R_s$ and $R_{ct}$ obtained from the fitting are 2.22 and 43.67 Ω, which ensure the effective ionic/electronic transport at the electrode-electrolyte interface and good ohmic contact with the current collector. The better charge-kinetics of designed ASC is the result of binder-free electrode fabrication which reduced the contact resistance. The pronounced performance of the designed ASC device is attributed to (i) large accessible surface area for the electrolyte by non-agglomerated vertical network electrode, (ii) excellent conductive pathways, (iii) superhydrophilicity of vertical TMO-graphene composite and (iv) superior electrochemical and mechanical stability.

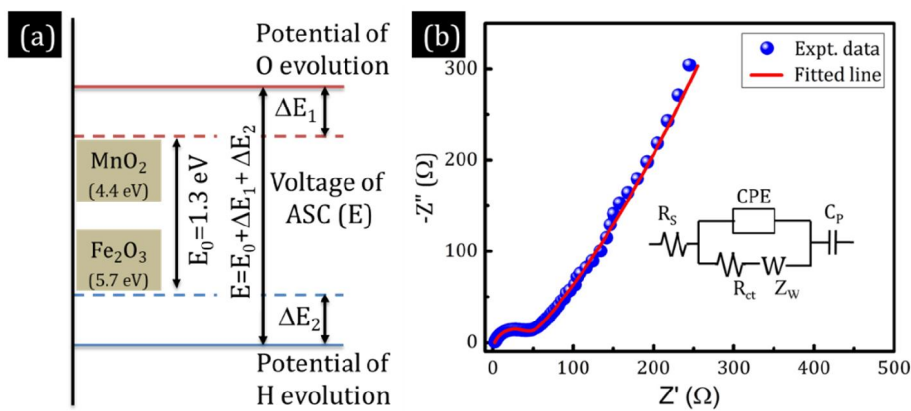

*Fig. 7: (a) Schematic of the stable potential window and (b) Nyquist plot of ASC*

## 4. Conclusion

A metal oxide-graphene hybrid electrode was successfully synthesized via feasible and facile techniques to fabricate an asymmetric Electrochemical capacitor device. The binder-free electrodes, $MnO_2$- vertical graphene as the positive and $Fe_2O_3$- vertical graphene as the negative electrode, separated by $NaClO_4$ soaked separator. Impressively, the fabricated $Fe_2O_3$-VGN//$NaClO_4$//$MnO_2$-VGN device showed the ability to operate with an extended window of 2.6 V. In addition to the double layer contribution, vertical graphene also supports the metal oxides structures and possess enhanced charge-storage capability. The fabricated ASC device showed a maximum capacitance of 76 mF/cm$^2$ with an energy density of 71 μWh/cm$^2$ and 79% capacitance retention over 12000 Charge-discharge cycles. This combination of hybrid electrode materials and electrolyte endeavor with the high-performance aqueous asymmetric device, showed its potential usage towards next-generation energy storage applications.

**Conflicts of interest**

There are no conflicts to declare.

**Acknowledgement**

This research was supported by Basic Science Research Program through the National Research Foundation of Korea (NRF) funded by the Ministry of Education (2017R1D1A1B03031989).

**Author contributions**

S G conceived the idea, carried out the growth, data analysis and manuscript preparation. S R P and G S conducted the VGN growth and Raman spectroscopy. All authors discussed the results, commented on the manuscript, and gave approval to the final version of the manuscript.